\documentclass[aip,graphicx,reprint]{revtex4-2}

\usepackage{graphicx}
\usepackage{xcolor}
\usepackage{caption}
\usepackage{subcaption}

\begin{document}


\title{Self-Assembly and Phase Behavior of Janus Rods: Competition Between Shape and Potential Anisotropy} 



\author{Jared A. Wood}
\affiliation{ARC Centre of Excellence in Exciton Science, School of Chemistry, University of Sydney, Sydney, New South Wales 2006, Australia}
\affiliation{The University of Sydney Nano Institute, University of Sydney, Sydney, New South Wales 2006, Australia}

\author{Laura Dal Compare}%
\affiliation{Dipartimento di Scienze Molecolari e Nanosistemi, 
Universit\`{a} Ca' Foscari di Venezia
Campus Scientifico, Edificio Alfa,
via Torino 155,30170 Venezia Mestre, Italy}

\author{Lillian Pearse}
\affiliation{School of Chemistry, University of Sydney, Sydney, New South Wales 2006, Australia}

\author{Alicia Schuitemaker}
\affiliation{ARC Centre of Excellence in Exciton Science, School of Chemistry, University of Sydney, Sydney, New South Wales 2006, Australia}

\author{Yawei Liu}
\affiliation{ARC Centre of Excellence in Exciton Science, School of Chemistry, University of Sydney, Sydney, New South Wales 2006, Australia}

\author{Toby Hudson}
\affiliation{School of Chemistry, University of Sydney, Sydney, New South Wales 2006, Australia}

\author{Achille Giacometti}%
\affiliation{Dipartimento di Scienze Molecolari e Nanosistemi, 
Universit\`{a} Ca' Foscari di Venezia
Campus Scientifico, Edificio Alfa,
via Torino 155,30170 Venezia Mestre, Italy}
\affiliation{European Centre for Living Technology (ECLT)
Ca' Bottacin, 3911 Dorsoduro Calle Crosera, 
30123 Venice, Italy}

\author{Asaph Widmer-Cooper}
\email{Asaph.Widmer-Cooper@sydney.edu.au}
\affiliation{ARC Centre of Excellence in Exciton Science, School of Chemistry, University of Sydney, Sydney, New South Wales 2006, Australia}\affiliation{The University of Sydney Nano Institute, University of Sydney, Sydney, New South Wales 2006, Australia}


\date{\today}

\begin{abstract}
We characterize the self-assembly and phase behavior of Janus rods over a broad range of temperatures and volume fractions, using Langevin dynamics simulations and free energy calculations. The Janus rods consist of a line of fused overlapping spheres that interact via a soft-core repulsive potential, with the addition of an attractive pseudo-square-well tail to a fraction of the spheres (the coverage) ranging from 5\% to 100\% of sites. Competition between the stability of liquid crystal phases originating from shape anisotropy and assembly driven by directional interactions gives rise to a rich polymorphism that depends on the coverage. At low density near the Boyle temperature, we observe the formation of spherical and tubular micelles at low coverage, while at higher coverage randomly oriented monolayers form as the attractive parts of the rods overlap. At higher density, bilayer structures appear and merge to form smectic and crystalline lamellar phases. All of these structures gradually become unstable as the temperature is increased until eventually regular nematic and smectic phases appear, consistent with the hard rod limit. Our results indicate that the intermediate regime where shape-entropic effects compete with anisotropic attractions provided by site specificity is rich in structural possibilities, and should help guide the design of rod-like colloids for specific applications.
\end{abstract}

\pacs{}

\maketitle 

\section{Introduction}

Janus particles, named after the two-faced Roman god, are particles with two chemically or physically distinct parts of their surface. They exhibit substantially more complex phase\cite{Li2013,Xu2015,Vissers2013,Otoole2017,OToole2017PhaseDumbells,Lattuada2011,Paiva2019,Sciortino2009} and dynamic\cite{Safaei2019} behavior than their single material analogues, driven by the different surface interactions of their two faces. For example, the dual nature of these particles means that they can exhibit similar self-assembly and phase behavior to molecular surfactants, forming lyotropic liquid crystal structures such as micelles\cite{Percebom2019} and able to stabilize Pickering emulsions\cite{Binks2001}.

Janus particles can be made in several ways. Surface coating part of a particle with a different material is one method. This typically involves trapping the particles and then depositing a material on one face\cite{Chaudhary2012, Zhang2017}. They can also be constructed from two different materials, either by growing a second material from a particle of one type\cite{Park2012}, or by growing two chemically incompatible materials from the same seed, such as a block co-polymer\cite{Lattuada2011, Percebom2019}.  

Spherical Janus particles are the most widely studied and represent a paradigmatic example of colloids whose phase behavior is dominated by directional interactions. On the other hand, a fluid formed by hard rod-like particles is a paradigmatic example of a colloidal suspension where the phase behaviour is dominated by shape anisotropy, giving rise to entropically stabilized liquid crystal structures \cite{Bolhuis1997,Lopes2021}. While a huge literature exists for these two cases, fewer studies have considered a combination of directional interactions and shape anisotropy. One example is provided by colloidal dumbbells/dimers \cite{Munao2013,Munao2014,Munao2015,Otoole2017,OToole2017PhaseDumbells} and ``Mickey-Mouse'' shaped colloids \cite{Wolters2015,Avvisati2015}, where the anistropic colloidal particles are formed by gluing together two or more differently-functionalized spherical colloids. 

Janus rods are harder to make, because their shape asymmetry requires extra control over where coating or growth of a second material takes place. Several groups, however, have overcome these limitations to make rods with attractive tips\cite{Chaudhary2012, Repula2019}, or Janus ellipsoids where the central axis is in the plane separating the two regions\cite{Shah2015}. Other possible routes are via cation exchange or synthesis of metal tips on semiconducting nanorods\cite{Sadtler2009,Mokari2005}. The self-assembly and phase behavior of such particles have also received some attention \cite{Liu2012,Xu2015,Tripathy2013,Chaudhary2014,Filippo2024}, but much less so than for Janus spheres and dumbbells.

In this work, we use computer simulations to characterize the self-assembly and phase behavior of Janus rods with different fractions of attractive sites ($\chi$) along their length, ranging from 5\% to 100\%. The rods can be regarded as very stiff diblock copolymers \cite{Doi2013}, such as cylindrical polymer brushes \cite{Verduzco2015,Mullner2016}, or inorganic rod-shaped colloids\cite{Mokari2005,Sadtler2009,Chaudhary2012}. In real systems, the dual nature of the two different moieties of the rods stems for instance from the hydrophobic/polar interactions of the different parts of the rods with the solvent, and are similar in spirit to conventional surfactants \cite{Doi2013}.

Our aims here are twofold. As the assembly and phase behavior depend on both the temperature and the rod concentration, we would like to identify the most interesting parts of the phase diagrams for different fractions of attractive sites (coverage). As we show, the assembly behavior changes with coverage, not only in terms of the relevant temperature ranges. Second, we would like to probe the low temperature limit for the low coverage case, where self-assembly driven by directional interactions strongly competes with the formation of liquid-crystalline phases driven by shape anisotropy. Our rods here are achiral. In this respect, this analysis is complementary to that of the companion paper on Janus helices \cite{DalCompare2023}, which is devoted to exploring the consequences of a microscopic chirality on the liquid crystal phases that form at relative high temperatures and concentrations.

\section{Model and Methods}
\label{sec:methods}
Reduced units are used throughout this paper, with the length scale set by $D$, the diameter of the spheres, and the energy scale set by $\varepsilon_0$. So $T^{*}=k_BT/\varepsilon_0$ and $P^{*}=PD^3/\varepsilon_0$ and the asterisk will be dropped hereafter.

\subsection{Models}
To model the nanorods, we used a rigid body composed of $n_s=20$ spheres evenly spaced over a length of $L/D=5$ sphere diameters\cite{Liu2019}. This forms rods with a tip-to-tip aspect ratio of 6, which allows us to connect with previous studies of fully repulsive and attractive rods\cite{Bolhuis1997,Gamez2017}. These spheres were divided between attractive and repulsive. Repulsive spheres interacted with all other spheres using a pseudo-hard-sphere Mie potential\cite{Mie1903}, truncated at the first zero ($r=1.0204$) \cite{Zeron2018, Liu2019}:
\begin{equation}
\label{sec1:eq1}
    U_{PHS}(r) = \left(\frac{n}{n-m}\right)\left(\frac{n}{m}\right)^{m/(n-m)}\epsilon_0\left[\left(\frac{\sigma}{r}\right)^n - \left(\frac{\sigma}{r}\right)^m\right]
\end{equation}
Here $n=49$, $m=50$, $\epsilon_0=1$, $\sigma=D$ and $r$ is the distance between the two spheres. 
The attractive spheres interacted with other attractive spheres using a pseudo-square-well potential \cite{Zeron2018}.
This potential is given by 
\begin{equation}
\label{sec1:eq2}
U_{PSW}(r)=\frac{\epsilon}{2}\left(\left(\frac{1}{r}\right)^{n}+\frac{1-\mathrm{e}^{-m(r-1)(r-\lambda)}}{1+\mathrm{e}^{-m(r-1)(r-\lambda)}}-1\right)
\end{equation}
which differentiates smoothly to 
\begin{equation}
\label{sec1:eq3}
F_{PSW}(r)=\frac{\epsilon n}{2}\left(\frac{1}{r}\right)^{n+1}-\frac{\epsilon m(2 r-\lambda-1) e^{-m(r-1)(r-\lambda)}}{\left(1+e^{-m(r-1)(r-\lambda)}\right)^{2}}
\end{equation}
Here $\lambda=1.5$ is the range of the square well, $\epsilon=\varepsilon_0/n_s^2=\varepsilon_0/400$, $m=50$ and $n=200$. $m$ and $n$ are free parameters used to tune the well shape. The choice of $\epsilon=\varepsilon_0/n_s^2$ is to reduce the dependence of the results on the choice of $n_s$ and to match the continuous line model\cite{Gamez2017}. Indeed, this choice allows us to reproduce the phase diagram for the fully attractive continuous model up to a minor shift of the boundaries (see Figure S1 in the supplementary material).

\begin{figure}
    \centering
    \includegraphics[width=90mm]{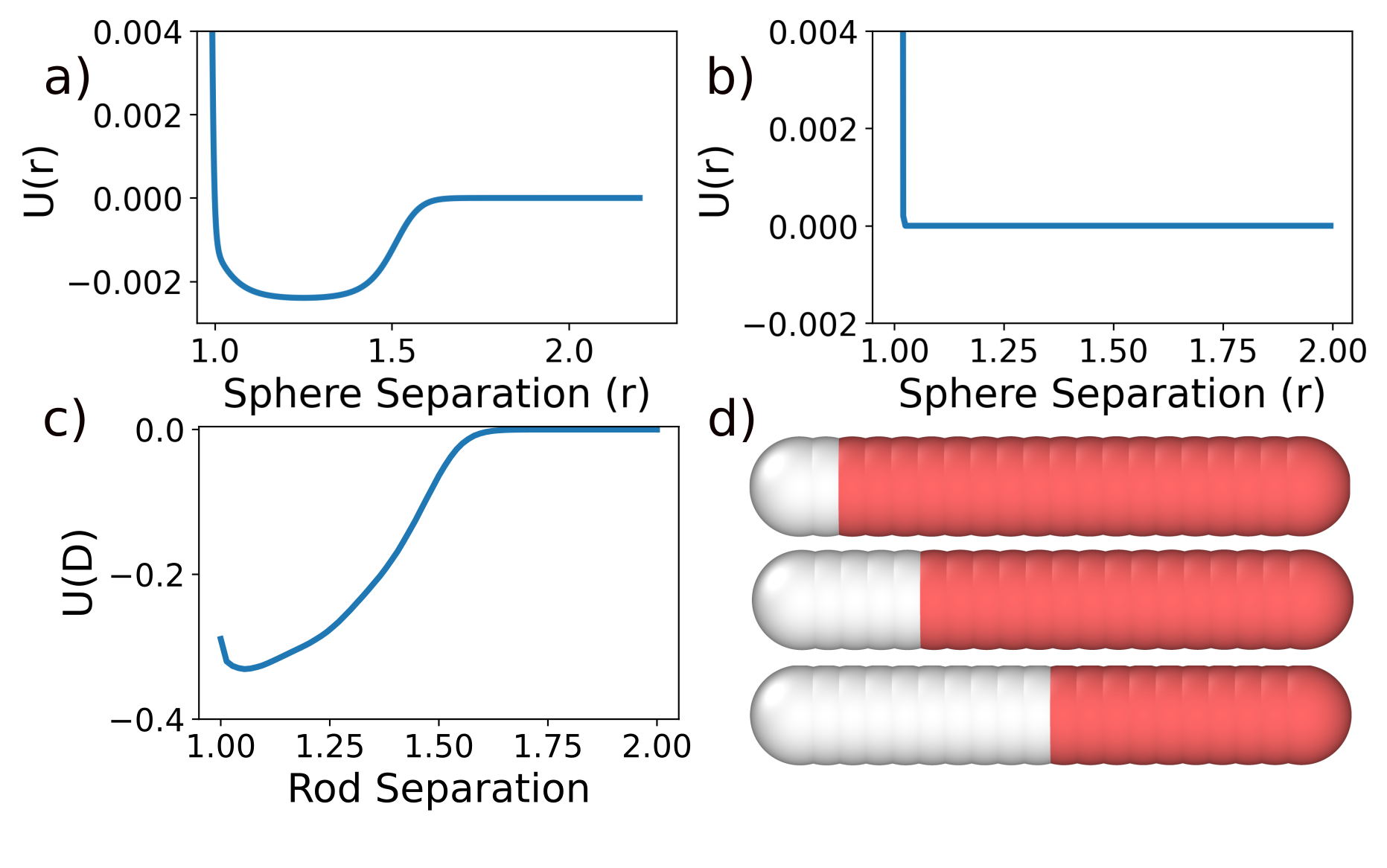}
    \caption{a) The pseudo-square-well interaction $U_{PSW}$ between two attractive spheres as a function of separation $r$. b) The repulsive Mie potential $U_{PHS}$ as a function of separation $r$. c) The total potential energy between two side-by-side parallel rods with $L=5$ composed of $n_s=20$ spheres as a function of core separation $D$. d) Examples of Janus rods with $\chi=0.1$, 0.25, and 0.5. Attractive spheres are white, and repulsive spheres are red.}
    \label{fig:potential}
\end{figure}

The form of the potentials, and the total potential between two parallel rods with $L=5$ approaching parallel is shown in Figure~\ref{fig:potential}.

\subsection{Langevin Dynamics Simulations}
All simulations with $L/D=5$ rods were performed using LAMMPS\cite{Plimpton1995} with a Langevin thermostat, to capture the basic components of Brownian particle dynamics in an implicit solvent. The mass of each sphere was set to $0.05$ so that the total mass of each rod was 1.

Constant volume runs (in the canonical ensemble) were used to study the self-assembly as a function of $\chi$ and $T$ at a rod volume fraction of 0.1 (Section~\ref{sec:low_density}). These systems consisted of 270 rods and were initialized by equilibration at sufficiently high temperature to form a disordered gas-like state. The temperature was subsequently reduced from 0.1 to 0.001 over $4\times10^7$ timesteps. Some constant temperature runs (at constant volume) and some constant pressure runs (at constant temperature) were also used, as indicated in the text and figure captions. To control the pressure, a Berendsen barostat was used.

Pressure ramps and constant pressure runs were then used to study the phase behavior of rods with $\chi=0.05$ and 0.1 in more detail (Section~\ref{sec:phase_behavior}). Pressure ramps were performed at temperatures near $T_B$ (see below), with the pressure changed at a rate of 0.001 per $8\times10^8$ timesteps. Decreasing pressure ramps were initialized from an ordered double bilayer structure (consisting of 4 layers in an ABAB crystal, with the attractive ends touching each other). Increasing pressure ramps were initialized from a disordered gas of 576 rods. Constant pressure runs at $T_B=0.75$ to 20 were used to construct the phase diagram for rods with $\chi=0.05$. These were equilibrated for 800 million timesteps at pressures ranging from $1\times10^{-7}$ to 0.32.

\subsection{Molecular Dynamics Simulations}

In order to make connection with our previous paper on Janus helices \cite{DalCompare2023} we also performed some Molecular Dynamics (MD) simulations at constant volume and temperature (NVT) for $N=4068$ rods using the software package LAMMPS \cite{Plimpton1995} employing the Nos\'e-Hoover thermostat \cite{HooverPRA1985} to keep the temperature fixed.
These rods were modeled in the same way as the helices in previous work using $n_s=15$ spheres distributed on a length $L/D=10$.
To mimic hard-core repulsion, we used the Weeks-Chandler-Anderson (WCA) potential~\cite{Weeks1971}

\begin{eqnarray}
\label{sec1:eq3}
U_{WCA}\left(r\right) &=& 4 \epsilon \left[\left(\frac{D}{r}\right)^{12}- \left(\frac{D}{r}\right)^{6} + \frac{1}{4} \right]
\end{eqnarray}
which vanishes for $r\ge \sqrt[6]{2} D$ and acts on the repulsive-repulsive and repulsive-attractive bead pairs. And for attractive-attractive bead pairs, we used the $U_{PSW}(r)$ (\ref{sec1:eq2}) potential described above.

\subsection{Umbrella Sampling}
The successive umbrella sampling method\cite{Virnau2004, OToole2017PhaseDumbells, Jungblut2008} was used to search for free energy barriers between different types of order formed by rods with low $\chi$ near $T_B$. These simulations were carried out using custom modifications (min and max keywords) to the \emph{fix gcmc} command in LAMMPS\cite{Wood2019}, which are now part of the stable release. It was necessary to use tabulated potentials for these simulations to avoid a known problem with how the energy is calculated\cite{Hafreager2017}, and to scale the temperature of inserted particles by 1.609 to maintain constant temperature, due to kinetic energy cancellation in the rigid body\cite{Lammps_authors2020}. Ten insertion/deletion moves were performed every 100 timesteps.

These grand-canonical simulations were run at constant volume in a cubic simulation box with side length 20, with the initial seed consisting of either a single rod or a micelle consisting of 20 rods.
The number of particles allowed in each simulation was restricted to 3 consecutive integers, though generally 2 dominated each simulation. Moves attempting to add or remove particles outside these bounds were rejected. Starting from the window 1-3, or 20-22 in the case of a pre-seeded run, we generated the initial state for subsequent simulations from each preceding one, with an overlap of 2 with the previous window. This process was repeated until a suitable initial state could not be generated within the simulation time. Each run was continued until the free energy curves generated from the last two deciles had converged.
The free energy profiles were reconstructed using naive addition of step-wise energy differences based on the relation
\begin{equation}
  \exp(-\beta\Delta \Omega_n) = \frac{p_{n}(n+1)+p_{n-1}(n+1)}{p_n(n)+p_{n-1}(n)}
\end{equation}
where $p_n(m)$ is the probability of having $m$ particles in the simulation with a minimum of $n$ particles and $\Delta \Omega_n$ is the grand canonical free energy difference between having $n$ and $n+1$ particles in the system. $p_{0}(1)=p_{0}(2)=0$.
The same number of samples from each window was used in the free energy calculation.

\subsection{Boyle Temperature}
We use the Boyle temperature $T_B$, the temperature at which the second virial coefficient ($B_2$) is 0, as a reference point for consistency between the different attractive fractions of the rods. The Boyle temperature was calculated by the method in Munao et. al.\cite{Munao2015} and Yethiraj and Hall\cite{Yethiraj1991}.
In this method, by generating a large number ($N_c$) of randomly oriented, independent configurations of two rods ($i,j$) in a box of volume $V$, $B_2$ can be approximately determined by
\begin{equation}
    B_2(\beta\epsilon)=-\frac{V \left \langle f_{ij}\right \rangle}{2N_c}
\end{equation}
Here $f_{ij}$ is the Mayer function between the two rods
\begin{equation}
    f_{ij}(\beta,\mathbf{r}) = \exp[-\beta U_{ij}(\mathbf{r})] - 1
\end{equation}
$U_{ij}(r)$ is the total potential energy of the two rods with relative displacement, and orientation $\mathbf{r}$. $\langle \ldots \rangle$ is the average over all displacements and orientations.
$T_B$ is then identified by finding $\beta$ such that $\langle f_{ij}(\beta,\mathbf{r}) \rangle = 0$ using a gradient descent method, and then converting to a temperature by $T = 1/(\beta \epsilon)$. 

The calculated Boyle temperatures for $L/D=5$ used in this paper are listed in Table \ref{tab:Boyle_temps}. Boyle temperatures for $L/D=10$ are reported in the companion paper on Janus helices \cite{DalCompare2023}.

\begin{table}[]
    \centering
    \begin{tabular}{|c|c|c|c|c|}
\hline
$\chi$ & $T_B$ & Cluster size & Cluster nematic\\
\hline
5\% & 0.00083 & 6.9 & 0.02\\
10\% & 0.0030 & 5.6 & 0.09\\
25\% & 0.013 & 13 & 0.89\\
50\% & 0.033 & 39 & 0.88\\
75\% & 0.052 & 135 & 0.94\\
100\% & 0.076 & 270 & 0.94\\

\hline

\end{tabular}
    \caption{Calculated Boyle temperatures ($T_B$) for Janus rods with aspect ratio $L/D=5$ and attractive fraction ($\chi$). Also listed are the average cluster size and average cluster nematic for aggregates formed upon cooling to $T=0.001$ at a volume fraction of 0.1.}
    \label{tab:Boyle_temps}
\end{table}
\subsection{Cluster Identification}
Clusters of rods were defined by rods that felt substantial attraction to each other. Two rods were considered part of the same cluster if the distance between their attractive spheres was less than 1.5, i.e. within the range of the attractive interaction.

\subsection{Weighted Average Cluster Nematic Order Parameter}
To measure the alignment in the system, we calculated the average nematic order parameter $(S_i = \sum_{j=1}^{n_a}\left(3|{\hat{\textbf{u}}_a}\cdot {\hat{\textbf{u}}_b}|^2 - 1\right)/{2n_a}$ for each cluster in the system, where $n_a$ is the number of rods in cluster $a$, and $\hat{\textbf{u}}_a$ and $\hat{\textbf{u}}_b$ are the unit orientational vectors of rods $a$ and $b$. To calculate an average for the whole system, we used a weighted average over all clusters, weighted by the size of each cluster. A more general procedure, used in the companion paper on Janus helices \cite{DalCompare2023}, is based on the evaluation of eigenvalues and eigenvectors of the Veilliard-Baron tensor. This procedure is here unnecessary as a detailed analysis of the liquid crystal phases is not the main issue of the present study.

\section{Results and Discussion}
The phase behaviour of Janus rods is very rich and depends on temperature, volume fraction, and coverage (the fraction of attractive sites). Our analysis highlights two important regimes. First, the effect of coverage at low volume fraction and temperature. Here, a progressive increase of the coverage drives the system to self-assemble into different morphologies driven by competition between strong directional short range interactions and excluded volume. This will be discussed in Section \ref{sec:low_density}, where connection will also be made with results of the companion paper on Janus helices \cite{DalCompare2023}. The second regime that we focus on is the phase behaviour over the full temperature-volume fraction plane at low coverage ($5-10\%$ in the present case). This allows us to make connection with experiments and numerical simulations from a previous study on viruses \cite{Repula2019}. This will be discussed in Section \ref{sec:phase_behavior}.
\subsection{Self-Assembly Behavior of Rods with $L/D=5$ at Low Volume Fraction}
\label{sec:low_density}
To determine the influence of the fraction of attractive sites $\chi$ on the types of structures that are stabilized by attractive interactions between Janus rods, we started by surveying the self-assembly behavior of the $L/D=5$ rods at a volume fraction of 0.1. This involved the use of cooling runs from above the Boyle temperature to below it, as well as some runs at constant temperature (see Methods section for details).

\subsubsection{Rods with $\chi = 5\%$ and $\chi = 10\%$}
\begin{figure}
    \centering
    \includegraphics[width=85mm]{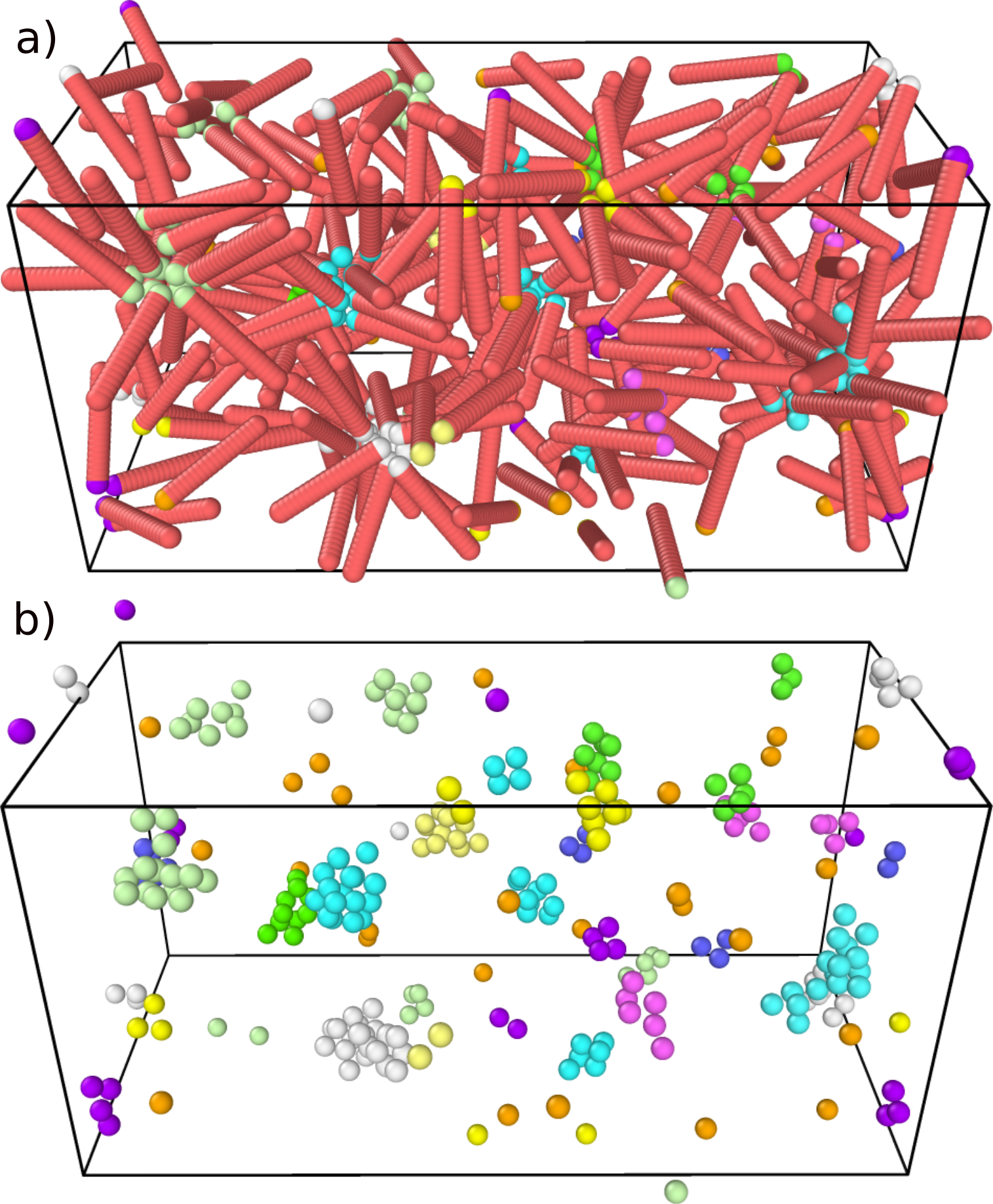}
    \caption{a-b) Final frame at temperature $T=0.0005$ from a constant temperature simulation of rods with $\chi = 5\%$, showing micelles with a) and without b) the repulsive portion of the rod. Here red beads are repulsive and other colored beads are attractive, with the color of the bead on the tip of the rod indicating which cluster the rod belongs to. Rods not in a cluster have orange tips. The rod volume fraction is 0.1.}
    \label{fig:tramp5}
\end{figure}
\begin{figure}
    \centering
    \includegraphics[width=85mm]{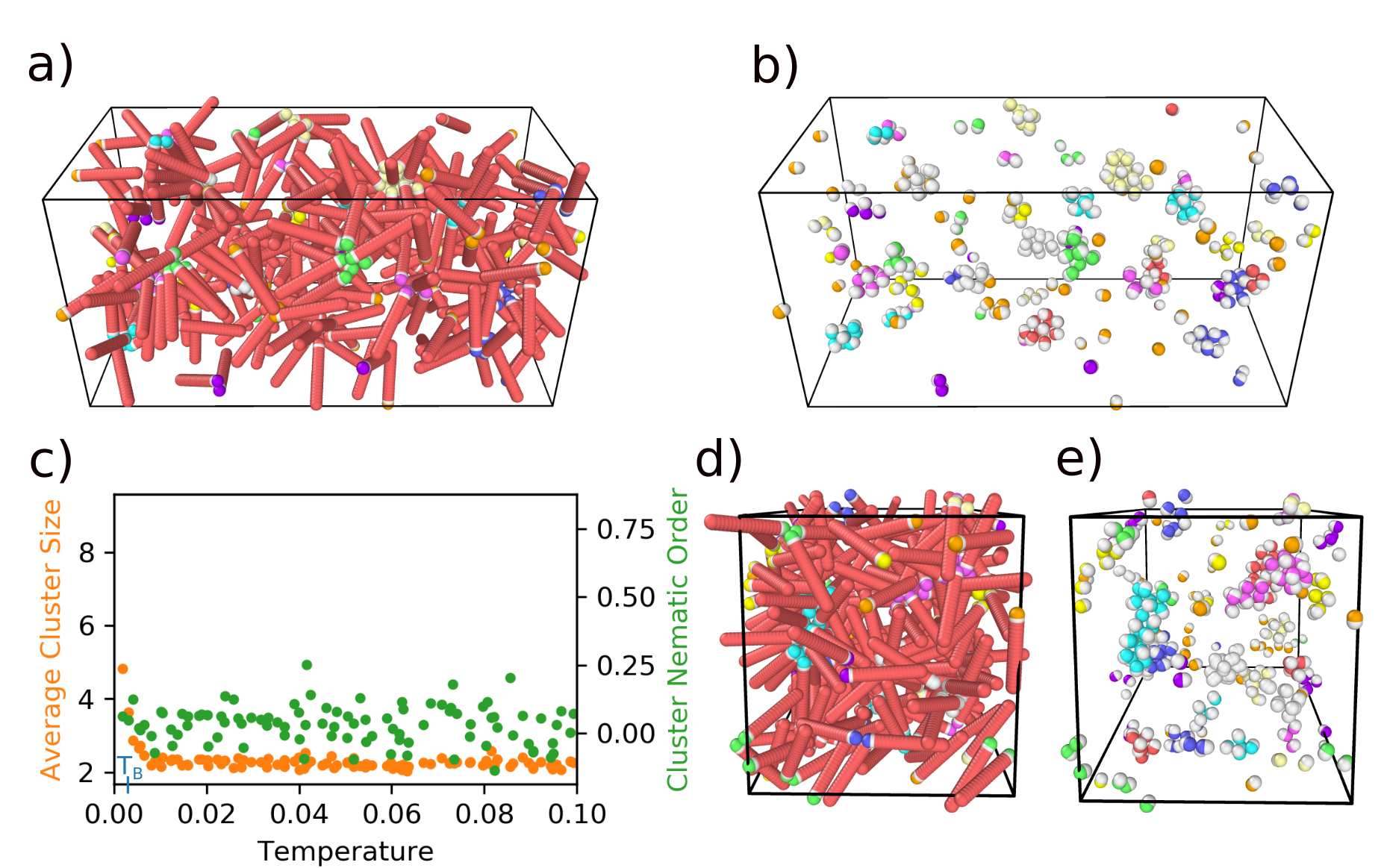}
    \caption{a-b) Final frame at temperature $T=0.001$ from a constant temperature simulation of rods with $\chi = 10\%$, showing micelles with a) and without b) the repulsive portion of the rod. Red beads are repulsive, white and other colored beads are attractive, with the color of the bead on the tip of the rod indicating which cluster the rod belongs to. Rods not in a cluster have orange tips. The rod volume fraction is 0.1. c) Average cluster size and average cluster nematic order parameter throughout a cooling run at the same volume fraction. d-e) Snapshot from a constant pressure run at $P=0.3$ and $T=0.75T_B$, showing the formation of tubular micelles (e.g. the teal cluster) with d) and without e) the repulsive portion of the rod.}
    \label{fig:tramp10}
\end{figure}

Upon cooling, the 5\% attractive rods remained isotropically dispersed until a temperature of about 0.002, roughly 2.5 times the corresponding Boyle temperature $T_B=0.00083$. Below this temperature, the attractive ends started to aggregate and micelle-like clusters of rods formed, with the attractive tips of the rods aggregating together and the repulsive part of the rods pointing outwards (Figure \ref{fig:tramp5}), similar to the micellar structures observed experimentally for Janus matchsticks \cite{Chaudhary2012}. These micelles were not uniform in size, and the orientation of the rods changed freely. Similar results were obtained for the 10\% attractive rods, as shown in Figure \ref{fig:tramp10}. For most of the cooling run, the rods remained isotropically dispersed, with micelles forming again near $T_B = 0.003$, as characterized by the increase in average cluster size without an associated increase in the cluster nematic order (Figure \ref{fig:tramp10}c and Table~\ref{tab:Boyle_temps}). Elongated or tubular micelles were also observed at slightly higher rod density during a constant pressure run at $P=0.3$ and $T=0.75T_B$ (Figure \ref{fig:tramp10}d-e). 

Note that $\chi=5-10\%$ roughly corresponds to a single attractive site in the companion paper on Janus helices \cite{DalCompare2023}. Below the Boyle temperature, we expect that those particles will also form micelles at low volume fraction, and indeed we show this below (in Section~\ref{sec:comparison}) for rods with the same contour length as the helices, i.e. $L/D=10$. At higher volume fraction, stable liquid crystal phases are formed by both Janus rods and helices. Hence, an intermediate regime when these two phases compete ought to exist, and this is further discussed in Section \ref{sec:phase_behavior}. 

\subsubsection{Rods with $\chi = 25\%$}
\begin{figure}
    \centering
    \includegraphics[width=85mm]{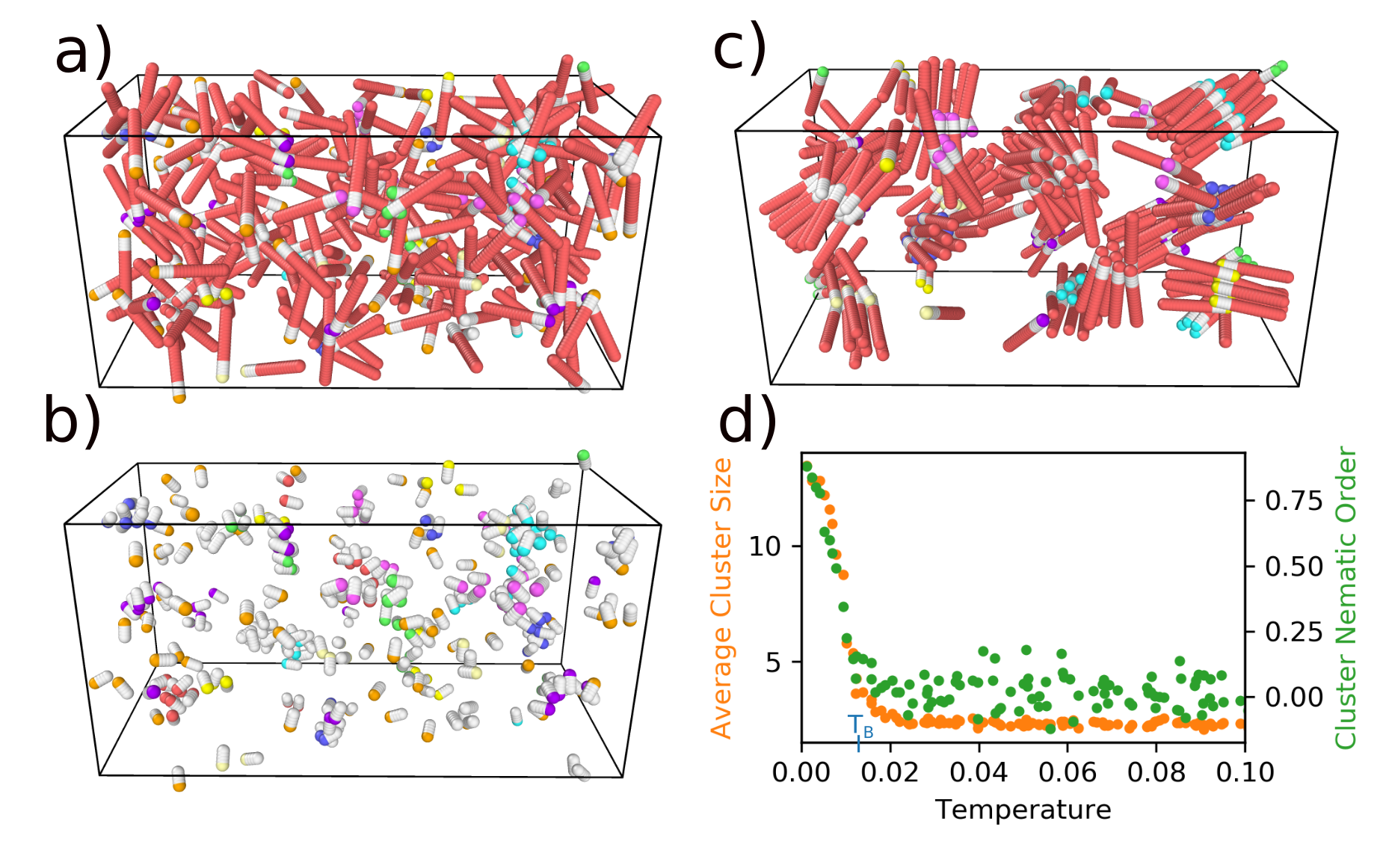}
    \caption{a-b) Snapshot taken near $T_B$ from a cooling simulation of rods with $\chi = 25\%$, showing the small clusters that initially form with a) and without b) the repulsive portion of the rod. Red beads are repulsive, white and other colored beads are attractive, with the color of the bead on the tip of the rod indicating which cluster the rod belongs to. Rods not in a cluster have orange tips. c) Final frame from the cooling run (at $T=0.001$), showing the randomly directed monolayers that eventually form. d) Average cluster size and average cluster nematic order parameter throughout the cooling run. The rod volume fraction is 0.1 in all cases.}
    \label{fig:tramp25}
\end{figure}

Upon cooling, the 25\% attractive rods remained dispersed until close to $T_B = 0.013$. At this point, the rods started to cluster, with the tips aggregating, much the same as seen for the 10\% attractive rods (Figure \ref{fig:tramp25}a-b). However, as the temperature further decreased, the rods in these clusters aligned with each other, and formed a randomly directed monolayer, with the attractive portions of the rods forming a single layer and the repulsive portions pointing randomly to either side, perpendicular to the layer (Figure \ref{fig:tramp25}c). This transition is characterized by a rapid increase in the cluster nematic order as the average cluster size increases (Figure \ref{fig:tramp25}d and Table~\ref{tab:Boyle_temps}).

Unlike regular monolayers, these layers are unable to pack together to form a dense multi-layered structure due to the random and sparse nature of the repulsive bodies on either surface. This generates sparse assemblies, as much of the space around the tail ends is empty, but inaccessible to other rods or clusters. This could frustrate the formation of more ordered bilayer structures as the density is increased, e.g. during solvent evaporation.

\subsubsection{Rods with $\chi = 50\%$ and $\chi = 75\%$}
\begin{figure}
    \centering
    \includegraphics[width=85mm]{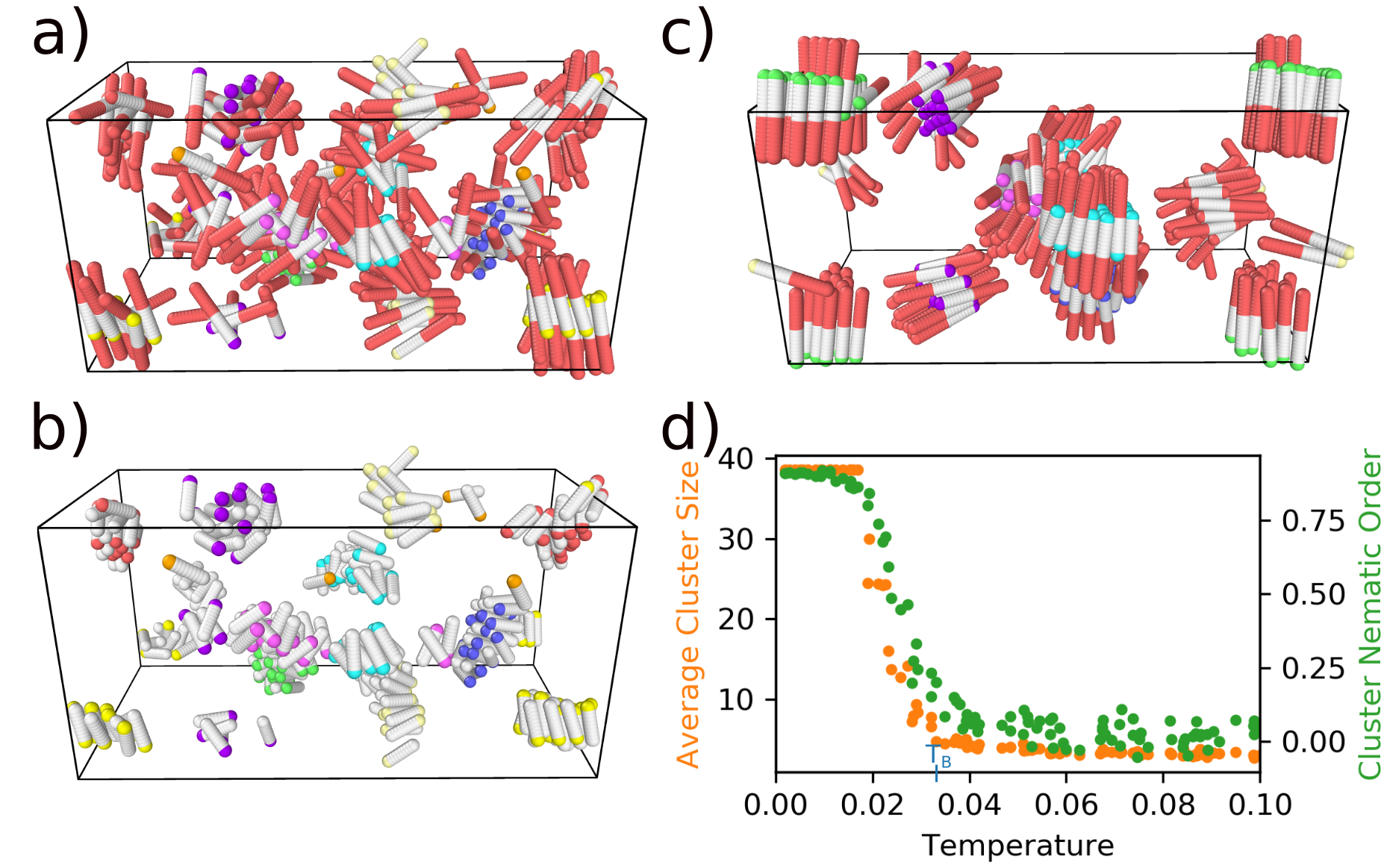}
    \caption{a-b) Snapshot taken near $T_B$ from a cooling simulation of rods with $\chi = 50\%$, showing the slightly twisted clusters formed with a) and without b) the repulsive portion of the rod. Red beads are repulsive, white and other colored beads are attractive, with the color of the bead on the tip of the rod indicating which cluster the rod belongs to. Rods not in a cluster have orange tips. c) Final configuration at the end of the temperature ramp (at $T=0.001$) showing the fully aligned randomly directed monolayer. d) Average cluster size and average cluster nematic order parameter throughout the cooling run. The rod volume fraction is 0.1 in all cases.}
    \label{fig:tramp50}
\end{figure}

When the 50\% attractive rods started ordering during the cooling run, they immediately formed randomly directed monolayers, with no micelle-like structures observed along the way (Figure \ref{fig:tramp50}. This is presumably because the greater length of the attractive segments biased their orientation parallel to each other more strongly. Surprisingly the assemblies that formed were not initially entirely parallel. Instead there was a twisting of the clusters at higher temperatures. Rather than aligning perfectly, the rods around the edge spontaneously tilted in unison, generating a chiral twist in the platelet, similar to that observed for smectic-like monolayer assemblies of long rod-shaped viruses\cite{Gibaud2012} and of model rods with $L/D=10$ \cite{Liu2022}. As we lowered the temperature further, this twist decreased until it disappeared, indicating that it is largely entropically stabilized in this case with no preferred handedness. This adds to a growing list of examples of spontaneous twisting observed in rod assemblies, with another recent example being the demonstration, using a simple Onsager-like theory, that achiral rods tend to twist in the nematic phase\cite{Revignas2023}.

Upon cooling, the 75\% attractive rods behaved largely the same as the 50\% attractive rods, forming twisted, randomly-oriented monolayers when they first aggregated (see Figure S2 in the supplementary information). Unlike the 50\% attractive clusters, these clusters maintained some of their twist down to the lowest tested temperature ($T=0.001$). This indicates that the twist in this case may also be at least partly energetically stabilized, similar to results recently obtained for fully attractive rods\cite{Liu2022} (see Section S6 of the Supplementary Information for that paper).

\subsubsection{Rods with $\chi = 100\%$}
\begin{figure}
    \centering
    \includegraphics[width=85mm]{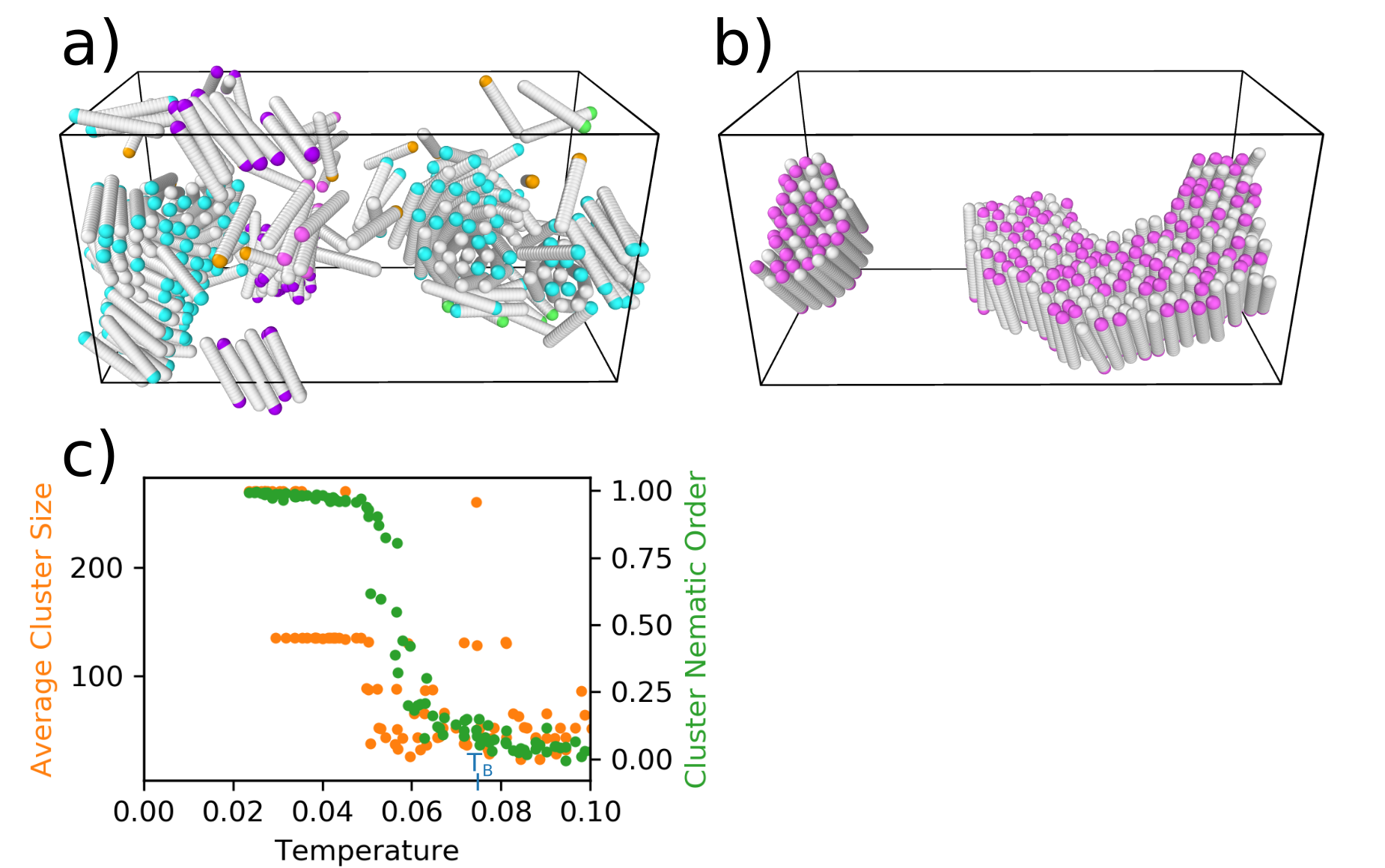}
    \caption{a-b) Snapshot taken near $T_B$ from a cooling simulation of rods with $\chi = 100\%$, showing the initial formation of twisted clusters. White and other colored beads are attractive, with the color of the bead on the tip of the rod indicating which cluster the rod belongs to. Rods not in a cluster have orange tips. b) Final configuration at the end of the temperature ramp (at $T=0.001$) showing the largely untwisted monolayer. c) Average cluster size and average cluster nematic order parameter throughout the cooling run. The rod volume fraction is 0.1 in all cases.}
    \label{fig:tramp100}
\end{figure}

Upon cooling, the 100\% attractive rods also aggregated into monolayers near $T_B$ (Figure \ref{fig:tramp100}). With no repulsive segment, however, these rods formed a single dense monolayer with no protrusions. At moderate temperatures, these monolayers also exhibited a twist, like that seen for the Janus rods with $\chi = 50\%$ and 75\% and for fully attractive rods with a larger range of attraction than in the present work (1.0 vs 0.5) \cite{Liu2022}. However, this twist largely disappeared at lower temperature, as the rods formed one large cluster. Compared to the behavior of the Janus rods, especially with $\chi = 75\%$, this might be due to the restoration of up-down symmetry and/or the larger cluster size. The relatively short range of attraction used in the present work (0.5) also favors the formation of crystalline monolayers, which are energetically costly to twist. The high degree of order within the monolayers is apparent in the radial distribution function for the rod midpoints at $T=0.001$ (shown in Figure S3), with the peak structure at $\chi = 100\%$ close to that of an ideal hexagonal packing in 2D. For the Janus rods, the order decreases from $\chi = 75\%$ to 25\% as the rod midpoints within the randomly oriented monolayers become offset from one another, before completely disappearing with the shift to micelles at $\chi = 10\%$ and 5\%.

\begin{figure*}[hbt]
  \centering
  \captionsetup{justification=raggedright,width=\linewidth}
  \begin{subfigure}[b]{0.45\linewidth}
   \includegraphics[trim=0 0 0 0,clip,width=\linewidth]{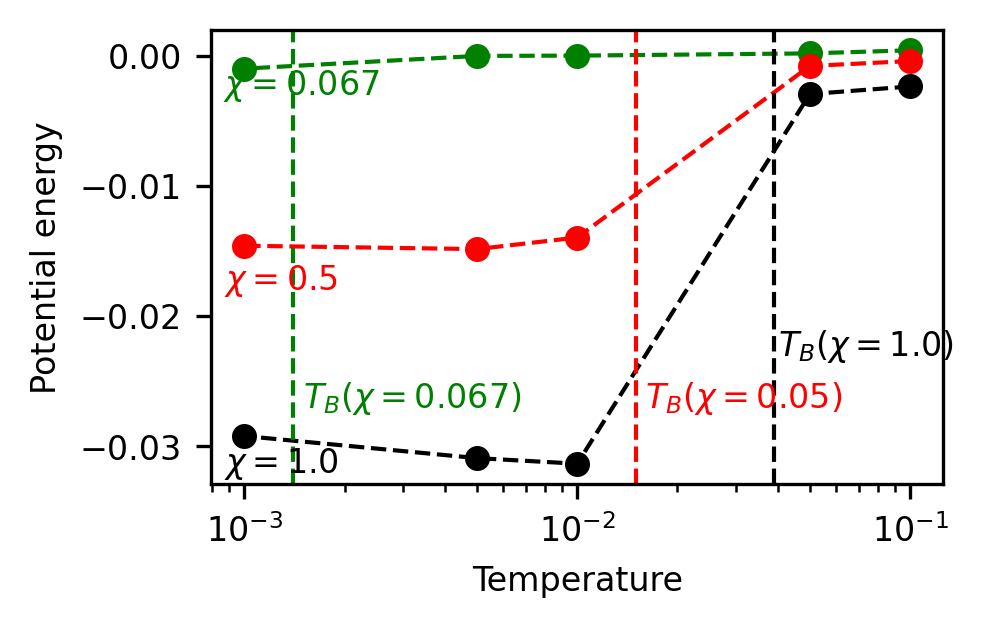}
   \caption{}\label{fig:fig7bisa}
  \end{subfigure}%
  \hfill
    \begin{subfigure}[b]{0.5\linewidth}
   \includegraphics[trim=10 10 10 0,clip,width=\linewidth]{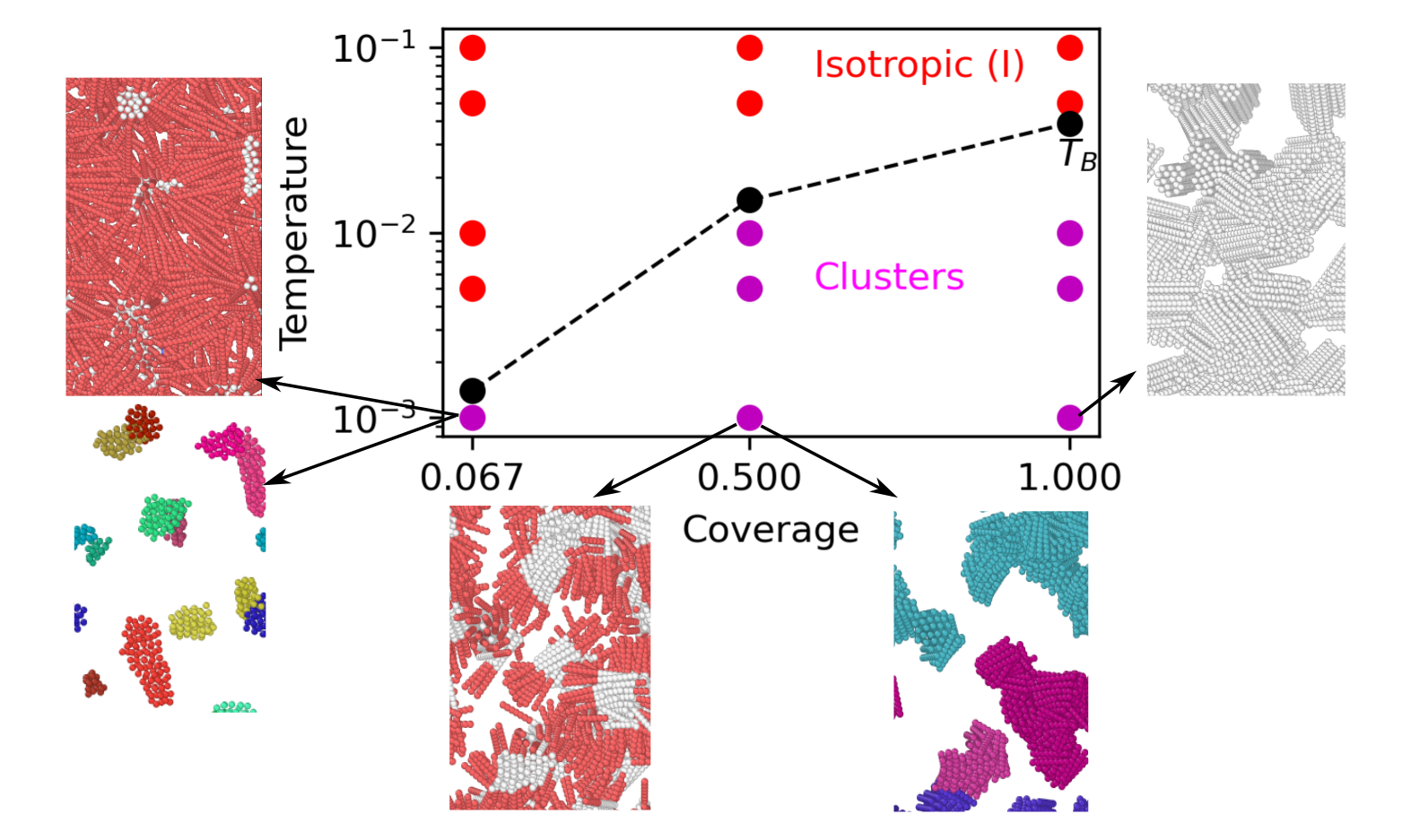}
   \caption{}\label{fig:fig7bisb}
  \end{subfigure}
  \caption{Self-assembly process for $L/D=10$ and volume fraction $0.1$. (a) Potential energy as a function of the reduced temperature $T^{*}=k_B T/\epsilon$ at different coverages $\chi$. Single attractive site $\chi=6.7\%=0.067$ (green), Janus half-fraction of attractive sites $\chi=50\%=0.5$(red), fully attractive rods $\chi=100\%=1.0$((black). Vertical dotted lines indicate the corresponding Boyle temperatures (color coded accordingly); (b) Corresponding phase diagram in the reduced temperature $T^{*}$- coverage $\chi$ plane. Insets report also representative snapshots in the different cases, including all parts of the rods (white attractive, magenta repulsive) or only the attractive rods color-coded according the specific cluster they are part of.}
  \label{fig:fig7bis}
\end{figure*}

\subsubsection{Comparison with the Self-Assembly Behaviour of Rods with $L/D=10$}
\label{sec:comparison}
To connect with our companion paper on the phase behavior of Janus helices \cite{DalCompare2023}, and a future paper on the self-assembly of those helices at low density, we present here results for Janus rods with the same contour length, i.e. $L/D=10$. And to connect with the results presented above, we keep the volume fraction at $\eta=0.1$ and consider three coverages: $\chi=6.7\%$ (corresponding to a single attractive bead), $\chi=50\%$ (corresponding to the Janus half fraction of attractive sites), and $\chi=100\%$ (corresponding to the case of fully attractive rods). Figure~\ref{fig:fig7bisa} shows the potential energy as a function of the reduced temperature for each coverage. In all cases, there is a decrease in the potential energy below the corresponding Boyle temperature (indicated with vertical dotted lines), marking the onset of the self-assembly process driven by the attractive sites. As expected, the decrease in energy is proportional to the coverage. The corresponding "phase diagram" is reported in Figure \ref{fig:fig7bisb} and shows that below the Boyle temperature (indicated by the black dotted line), the rods self-assemble into clusters whose structure depends upon the coverage in a similar manner to that described above for rods with $L/D=5$. The insets show that micelles form when $\chi=6.7\%$, randomly oriented monolayers form when $\chi=50\%$, and crystalline monolayers form when $\chi=100\%$.

In comparison with our previous study of Janus helices above the Boyle temperature \cite{DalCompare2023}, where the phase behavior was largely determined by excluded volume, the present results indicate that at low temperature and low density the phase behaviour is instead driven by the need to maximise the number of favorable contacts. That said, preliminary results for Janus helices at low temperature and density indicate that the chirality of the helices also provides a local twist that is not present for rods. A detailed analysis of this effect will be reported in future work.

\subsection{Phase Behavior of Rods with $L/D=5$ at Low $\chi$}
\label{sec:phase_behavior}
To further investigate how the phase behavior of Janus rods is affected by competition between entropic shape effects and potential anisotropy, we focused on rods with $\chi = 5\%$ and 10\%. As discussed above, these rods assembled into micellar structures at low density and temperature due to their potential anisotropy. In contrast, we know that they will form nematic and smectic phases at higher temperatures and densities. To study the transition between these regimes, we initially used pressure ramps to roughly map out the phase behavior near the Boyle temperature and far above it. We then focused on the case of $\chi = 5\%$ and mapped out the phase diagram in detail using constant pressure runs over a wide range of temperatures and pressures. Finally, we used cluster analysis and free energy calculations in an attempt to identify barriers between the low-temperature phases. Full details are provided in the Methods section.

\subsubsection{Pressure Ramps}
Pressure ramps were run at temperatures ranging from $20T_B$ to $T_B$. The decreasing pressure ramps were started from a periodic bilayer hexagonal crystal, with the attractive tips in alternating layers pointing towards each other, which is the lowest potential energy configuration. The increasing pressure ramps were started from a partially equilibrated random gas configuration at very low pressure. The pressure was then gradually changed over the length of the simulation. By comparing the two pressure ramps, it is possible to determine when the runs are close to equilibrium and when kinetic effects, such as metastability associated with phase transitions, are at play. As similar results were obtained for rods with $\chi = 5\%$ and 10\%, we discuss these cases together.

\begin{figure}
    \centering
    \includegraphics[width=80mm]{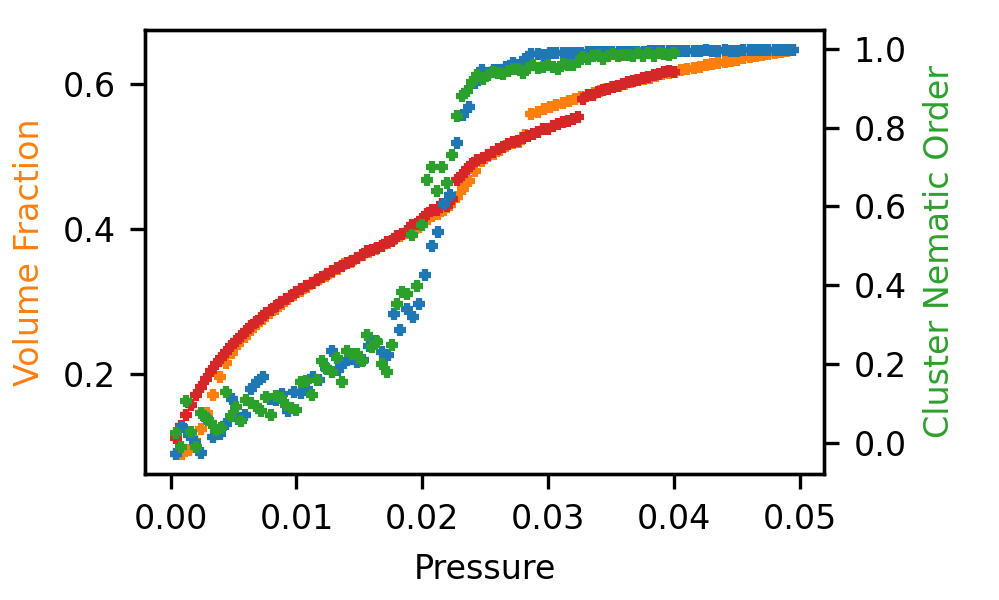}
    \caption{Results from pressure ramps at $20T_B$ for rods with $\chi=5\%$. The volume fraction (orange, red) and average cluster nematic order (blue, green) are plotted for increasing (orange, blue) and decreasing (red, green) pressure ramps.} 
    \label{fig:pramp 5-20}
\end{figure}

Well above the Boyle temperature, the phase behavior is dominated by entropic shape effects. At $20T_B$, we observe all of the phases that are formed by hard rods with the same aspect ratio \cite{Bolhuis1997}, i.e. isotropic, nematic, smectic and crystalline phases (Figures \ref{fig:pramp 5-20} and S4)). On decreasing the pressure (red and green data points), the initially crystalline structure transforms into a smectic phase and then a nematic phase (both observed as sudden decreases in the volume fraction), before transforming into an isotropic fluid phase. The presence of the nematic phase is confirmed by a small range of pressures below the smectic-nematic transition over which the nematic order parameter is elevated. On increasing the pressure, the inverse sequence of phases is observed, with some hysteresis of the nematic-smectic and smectic-crystal transitions. The smectic and crystal phases that are thus formed lack the bilayer structure of the lowest energy state, confirming that the potential anisotropy is insignificant at this temperature.

\begin{figure}
    \centering
    \includegraphics[width=80mm]{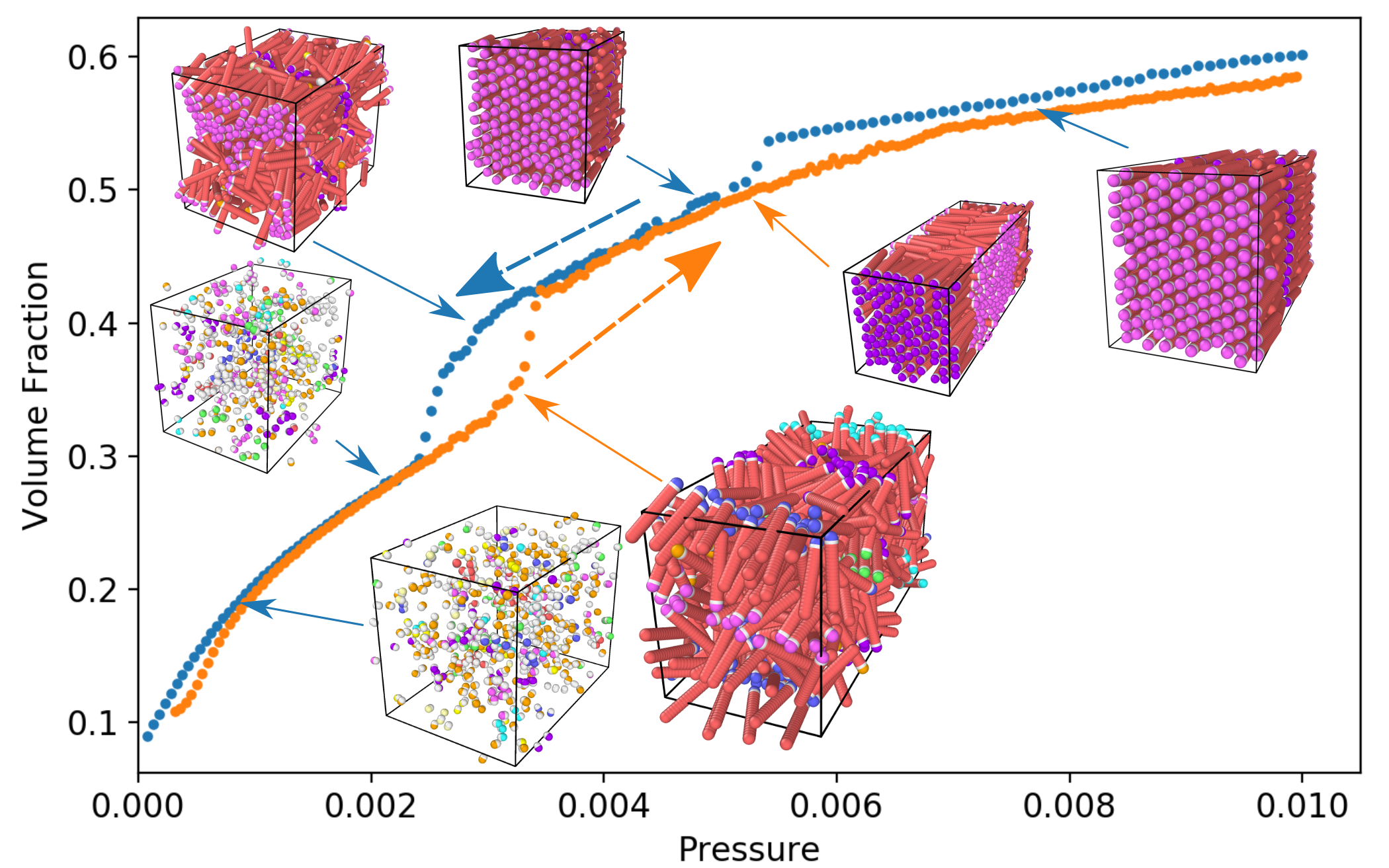}
    \caption{Pressure ramp results for rods with $\chi = 10\%$ at a temperature of $2T_B$. The insets are snapshots highlighting the different structures formed. Blue data points are from a decreasing pressure ramp, starting from two crystal bilayers. Orange data points are from an increasing pressure ramp starting from a disordered gas.}
    \label{fig:pramp 10-2}
\end{figure}

\begin{figure}
    \centering
    \includegraphics[width=80mm]{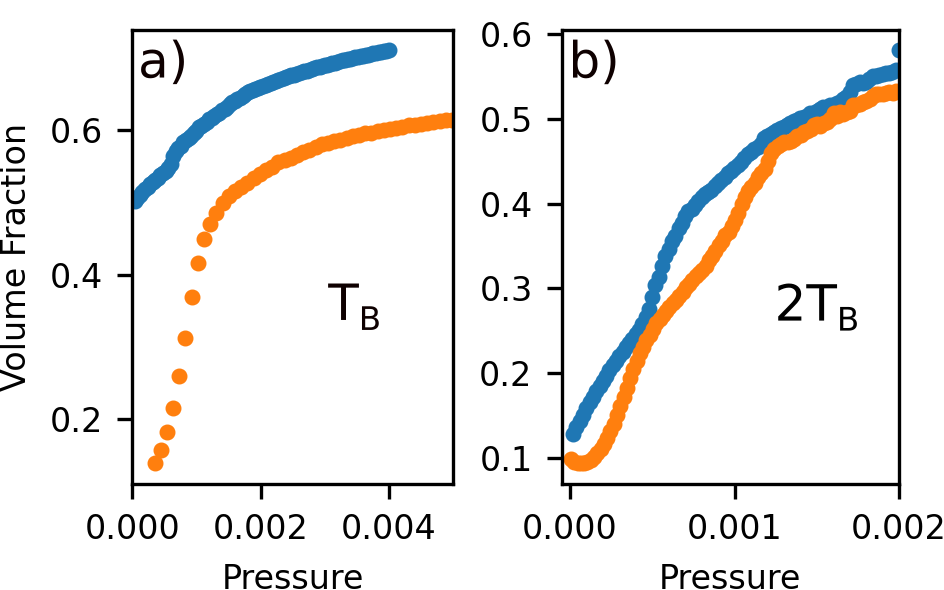}
    \caption{Volume Fraction for increasing (orange) and decreasing (blue) pressure ramps for $\chi=5\%$ rods at temperatures of a) $T_{B}$ and b) $2T_{B}$.} 
    \label{fig:pramp 5-2}
\end{figure}

At twice the Boyle temperature, the phase behavior is already strongly influenced by the potential anisotropy (Figure \ref{fig:pramp 10-2}). Upon increasing the pressure, the initially isotropic fluid appears to form densely packed tubular micelles, which then transform into a bilayer structure (with stacking fault), where the attractive tips point towards each other. Starting from a perfect bilayer crystal and decreasing the pressure, the system first transforms into a bilayer smectic, before transforming into an isotropic fluid. The absence of tubular micelles upon decompression is likely related to the large hysteresis observed for the latter transition. Similar results are obtained for $\chi = 5\%$ (Figure \ref{fig:pramp 5-2}b), with the phase transitions shifted to lower pressure. This behavior is consistent with that found in Ref. \cite{Repula2019}, where it was noted that the smectic phase is stabilized against the nematic phase by a strong localized tip attraction (see their Figure 2).

At the Boyle temperature, for both values of $\chi$, we observe substantial hysteresis in the pressure ramps (Figures \ref{fig:pramp 5-2}a and S5). Upon decompression, the initial bilayer structure remains intact down to zero pressure, with the only defect being gaps between the repulsive rod ends at very low pressure for $\chi = 5\%$. In contrast, upon compression the initially random gas of rods appears to form spherical micelles, then tubular micelles, before transforming into a bilayer structure. It is therefore clear that at the Boyle temperature the phase behavior is dominated by the potential anisotropy. These runs, however, provide little reliable information about the phase boundaries near $T_B$ due to the substantial hysteresis.

\subsubsection{Phase Diagram}
\begin{figure}
    \centering
    \includegraphics[width=88mm]{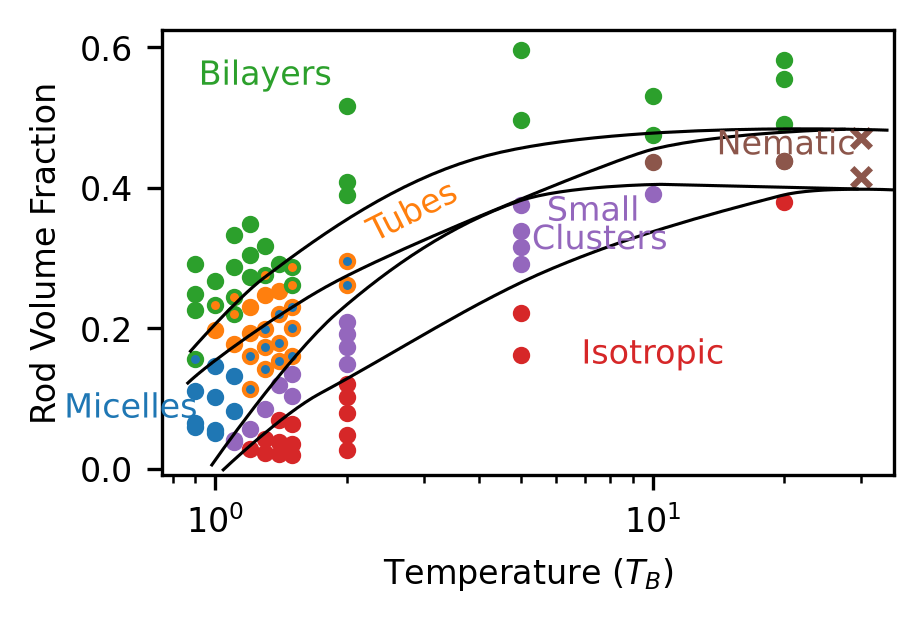}
    \caption{Phase diagram for rods with $\chi=5\%$ in the volume fraction-temperature plane. Temperatures are measured in units of the Boyle temperature $T_B$, with the crosses at $30T_B$ marking the boundaries of the nematic phase in the high temperature limit \cite{Bolhuis1997,Gamez2017}.  Data at 20$T_B$ was taken from a decompression ramp. All other data was taken from equilibrated simulations at constant pressure and temperature. Points with different center and outline colors indicate co-existence between different types of order.} 
    \label{fig:5phase}
\end{figure}

To determine where the phase boundaries are located near $T_B$, and how the phase behavior changes as a function of temperature, we used runs at constant pressure and temperature to carefully map out the phase diagram for the rods with $\chi = 5\%$. A phase diagram generated from these simulations is shown in Figure \ref{fig:5phase} in the temperature-volume fraction plane.

Close to $T_B$ we observe micelles forming at low rod volume fraction, which gradually merge to form tubular micelles and then eventually bilayer clusters upon increasing the volume fraction. Above $T_B$, the stable phase at low volume fraction is the isotropic one, with both positional and orientational disorder. Here upon increasing the volume fraction, the rods aggregate into small clusters, tubes and eventually dense smectic and crystalline bilayers. As the temperature increases, the phase boundaries shift to higher volume fraction, until the nematic phase appears around $10T_B$ at a volume fraction around 0.45. 

When assigning phases, we regarded all systems with an average cluster size of $\leq1.7$ to be isotropic. Micelles were identified as near-spherical aggregates with an average cluster size of $>3$, and were only observed for systems with $T\leq2T_B$. At higher temperatures, the high volume fraction required to form clusters meant that the clusters were highly variable in size and shape, and thus could not be considered as micelles. Tubes were distinguished from micelles by having one dimension more than double the size of the other two dimensions when examining only the attractive spheres, not the repulsive ones. Bilayer clusters were distinguished from tubes by their high nematic order, while the nematic phase had a high nematic order parameter but a low average cluster size. For simplicity, we have not attempted to distinguish between smectic and crystalline bilayers, but our pressure ramps locate that boundary around a volume fraction of 0.5 at $2T_B$ and slightly higher at $20T_B$.

Though we see micelles form near $T_B$, it is difficult to identify a sharp phase boundary between spherical and tubular micelles, with clusters of different sizes and shapes co-existing in many of the constant pressure simulations (indicated by data points with a blue center and orange outline). Such coexistence between spherical and elongated micelles was also observed in a recent simulation study of similar end-patchy Janus rods, with other interaction parameters favoring more monodisperse micelle shapes and sizes \cite{Filippo2024}.

\subsubsection{Cluster Size Distribution and Order}
\begin{figure}
    \centering
    \includegraphics[width=80mm]{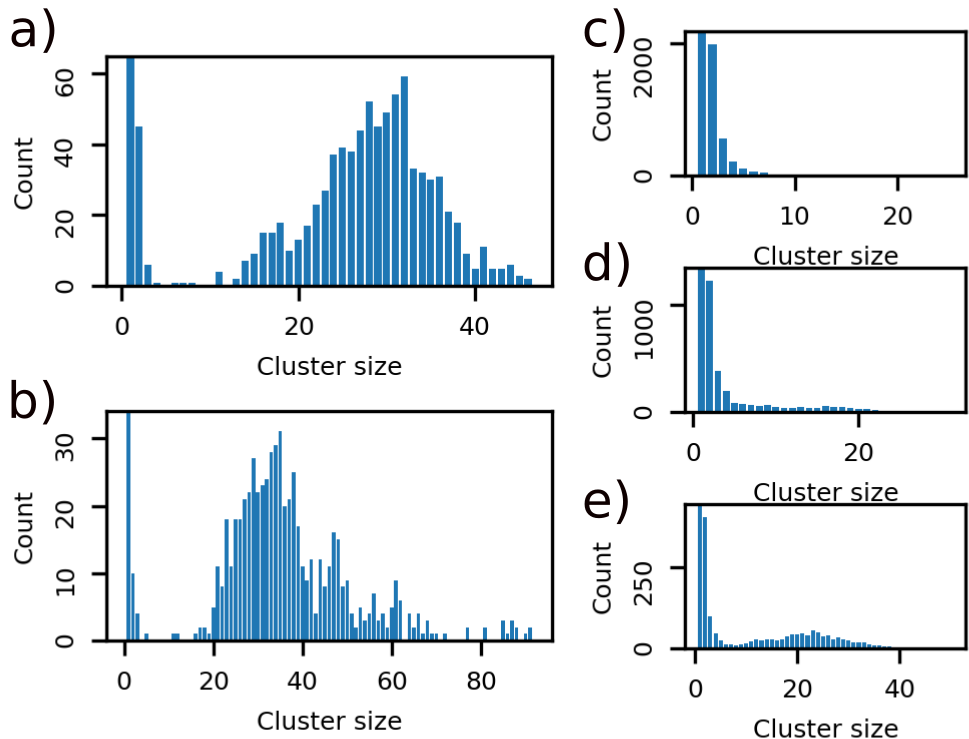}
    \caption{Cluster size distribution for rods with $\chi=5\%$ at $T_B$ (a,b) and $1.2T_B$ (c-e). The reduced pressures are (a) $1\times10^{-5}$, (b) $2.5\times10^{-5}$, (c) $1\times10^{-6}$, (d) $1\times10^{-5}$, and (e) $2.5\times10^{-5}$. These represent (a) micelles, (b) micelles and tubes, (c) isotropic, (d) small clusters, and (e) micelles. For legibility, the column indicating clusters of size 1 (free rods) has been truncated.} 
    \label{fig:5cluster_sizes}
\end{figure}

To illustrate the differences between the various low-density structures in our phase diagram, we have plotted some of their cluster size distributions (Figure~\ref{fig:5cluster_sizes}). All of the distributions exhibit an exponential decay near a size of 1, which is due to free rods that occasionally come into contact. This is the only structure seen in the isotropic systems, for example plot (c), while the systems we have identified as small clusters have a longer tail but no distinct second peak, such as plot (d). In contrast, systems that we have identified as containing micelles (a, b and e), have an additional broad distribution of cluster sizes centered around 20-30 rods, depending on the pressure and temperature. The appearance of this bimodal distribution indicates the critical micelle concentration, which in Figure \ref{sec:phase_behavior} corresponds to the phase boundary between the purple and blue/orange data points. As the elongated or tubular micelles start to form, additional larger clusters appear in the distribution outside the normal distribution present from the micelles, as shown in plot (b). 

The broad distribution of micelle sizes formed by these Janus rods is not typical, being roughly twice that reported in another coarse grained study of micelles \cite{Kopelevich2005}. Normal surfactant micelles have a narrower range of sizes, due to a strong preference by the molecules for a particular surface area/volume ratio\cite{Kopelevich2005, Schulz1991}. It therefore seems likely that the broad distribution of micelle sizes for our Janus rods is due to a weak preference for a particular surface area/volume ratio.

To further characterize the order in the systems analyzed in Figure~\ref{fig:5cluster_sizes}, we report several additional metrics in the Supplementary Material. These include the radial distributions functions (RDFs) for the rod midpoints (Figure S6) and the eccentricity distributions for the attractive beads in each cluster (Figure S8). The eccentricities were calculated as described in Ref. \cite{Murakami2021} and range from 0 for a perfect sphere to 1 for a highly asymmetric shape, with 0.66 corresponding to a major:minor axis ratio of 4:3. The micellar systems (a, b and e) exhibit a clear peak in the RDFs around $1.4D$, and a skew in the eccentricity distributions towards values greater than 0.5, both of which are lacking for the isotropic (c) and small cluster (d) systems. Consistent with our identification of elongated micelles in (b), this system shows the presence of the most asymmetric clusters with eccentricity values up to 0.95, corresponding to a major:minor axis ratio near 2:1. The structure factors for the rod midpoints (Figure S7), show the development of a peak at low wavevector as the rods start to aggregate, with the peak strongest for the elongated micelles (b) and absent for the isotropic case (c). This peak is characteristic of intermediate range order and can be caused either by clustering (as in this work) or by long-range repulsion\cite{Godfrin2014}.

\subsubsection{Transitions Between Low-Density Phases}
\begin{figure}
    \centering
    \includegraphics[width=80mm]{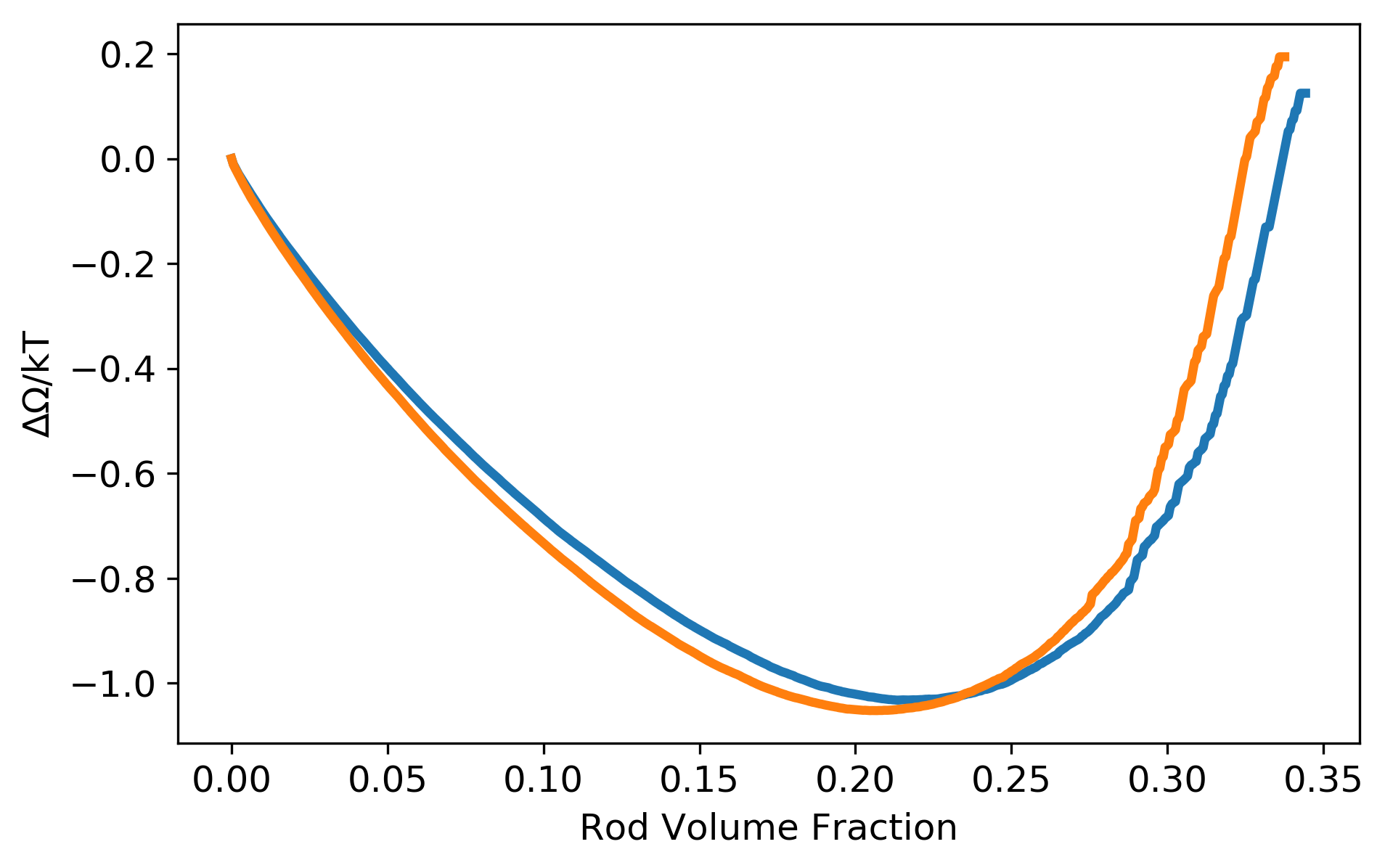}
    \caption{Grand canonical free energy change as a function of the rod volume fraction for rods with $\chi=5\%$ at $T_B$ (blue) and $1.1T_B$ (orange). $\mu=0$ for both simulations.} 
    \label{fig:5gcmc}
\end{figure}

In this section, we examine the nature of the transitions between the low density phases, i.e. the isotropic phase, the spherical micelles, and the tubular micelles. While one can quantitatively distinguish between them using metrics based on cluster size and order, the transitions between the different types of order are structurally much smoother than occurs, for example, at the isotropic-nematic and nematic-smectic phase boundaries. To determine whether there is a characteristic energetic change during these transitions, we used successive umbrella sampling to search for discontinuities in the free energy as a function of the rod volume fraction near the Boyle temperature, where the transitions are the most pronounced. The resulting grand canonical free energy change, shown in Figure \ref{fig:5gcmc}, is smooth and indicates that there are no free energy barriers associated with the various phase transitions. The volume fraction covered extends from one rod in the box up to the formation of large tubular networks and some discrete bilayers. Beyond a volume fraction of $0.25$ the curve becomes noisier due to a low move acceptance rate, but remains smooth with no discontinuities that may indicate a first-order transition. Similar results are obtained for rods with $\chi=10\%$, as shown in Figure S9.

The chemical potential in these simulations was tuned by trial and error to put the free energy minimum within the volume fraction range reachable by the simulations. Changing $\mu$ changed the position and depth of the minimum, but was unimportant aside from ensuring a good acceptance rate in the particle addition and removal moves. It does not change the magnitude of any free energy barriers that may exist between thermodynamically distinct phases.

\section{Conclusions}
We studied the self-assembly and phase behaviour of a system of Janus rods with aspect ratio $L/D=5$, where a fraction of the rod sites (coverage) have a pseudo-square-well attractive tail. This aspect ratio was chosen to connect with previous studies of hard and fully attractive rods\cite{Bolhuis1997,Gamez2017}. The rods were investigated using Langevin dynamics simulations and umbrella sampling, primarily at temperatures close to or below the corresponding Boyle temperatures, i.e. when the aggregation process is predominantly driven by enthalpy. Two different regimes were probed. First we characterized the self-assembly behavior upon cooling at low volume fraction (0.1). At low coverage ($\chi=5-10\%$),  we observed the formation of micelles driven by attraction between the rod tips starting near the corresponding Boyle temperature. As the coverage increased to $\chi=25\%$, micelles were replaced by monolayers formed by the attractive portions of the rods, with the repulsive portions randomly oriented. This increases the rotational entropy without incurring an energetic penalty as long as the attractive regions overlap. This orientational degeneracy persists until the limit of fully attractive rods ($\chi=100\%$) is reached. Interestingly, at and above the Janus limit ($\chi \geq 50\%$), the aggregates tends to spontaneously twist to form chiral assemblies. The same tendency is observed for fully attractive rods ($\chi=100\%$ coverage) at intermediate temperature, but in this case the twist disappears upon further cooling.

To connect with previous work on Janus helices \cite{DalCompare2023}, we also characterized the self-assembly behavior of Janus rods with aspect ratio $L/D=10$ at a volume fraction of 0.1 (see Figure \ref{fig:fig7bis}). Similar to our results for rods with $L/D=5$, we observed self-assembly below the corresponding Boyle temperature into micelles, randomly oriented bilayers, or monolayers, depending on the coverage $\chi$.

Lastly, we characterized the phase behavior of rods with $L/D=5$ at low coverage ($\chi=5-10\%$) in the temperature-volume fraction plane. This was done to obtain insight into the transition between the low temperature low volume fraction regime, where micellar aggregates form as a result of competition between site-specific directional interactions and entropy favoring positional and orientational disorder of the rods, and the high temperature high volume fraction regime, where liquid crystal phases appear. What we found was a rich polymorphism, with tubular micelles, bilayer smectics and bilayer crystals appearing alongside spherical micelles and regular nematic and smectic phases. The high temperature limit was found to be consistent with previous studies of purely repulsive \cite{Bolhuis1997} and attractive rods \cite{Gamez2017}, as it should. In the micellar region, coexistence between monomers and clusters of 10--50 rods was observed, compatible with the existence of a critical micelle concentration that increases with temperature. Analysis using umbrella sampling found no free energy barrier between the various isotropic and micellar phases, and hence the transitions between these do not appear to be associated with any discontinuity in the thermodynamics.

Our results indicate that the intermediate regime where shape-entropic effects compete with anisotropic attractions provided by site specificity is rich in structural possibilities and deserves further study. This complexity should be even richer for the case of chiral Janus helices, which we have so far only studied at relatively high temperatures and densities in the companion paper \cite{DalCompare2023}.

\section*{supplementary material}
The supplementary material contains additional results for rods with $L/D=5$, including the phase diagram for fully attractive rods, radial distribution functions for all of the systems at low density, pressure ramps for rods with $\chi=10\%$ at $1-20T_B$, and free energy change as a function of the rod volume fraction for rods with $\chi=10\%$.

\begin{acknowledgments}
This work was supported by the Australian Research Council grants CE170100026 and FT140101061 (AWC and JW), the MIUR PRIN-COFIN2022  grant 2022JWAF7Y, the Erasmus + International Mobility Program, and the Eutopia COST Action CA17139 (AG). Computational resources were provided by the Sydney Informatics Hub, a Core Research Facility of the University of Sydney.
\end{acknowledgments}

\section*{Data Availability Statement}
The data that support the findings of this study are available from the corresponding author upon reasonable request.

\section*{Author declaration}
The authors have no conflicts to disclose.

\section*{Author contributions}
JAW: software, formal analysis, methodology, visualization, original draft preparation. LDC:  software, formal analysis, methodology, visualisation. LP: formal analysis, methodology, visualisation. AS: software, formal analysis, methodology, visualisation. YL: software, methodology. TH: conceptualization, review $\&$ editing. AG: conceptualization, methodology, funding acquisition, review $\&$ editing. AWC: conceptualization, formal analysis, methodology, funding acquisition, review $\&$ editing.

\section*{references}
\bibliography{refs}

\end{document}